\documentclass[aps,prd,preprint,superscriptaddress,tightenlines,%
nofootinbib]{revtex4}
\newcommand{\PRE}[1]{{#1}}   

\usepackage{amsmath}
\usepackage{amssymb}
\usepackage{color}
\usepackage{bm}
\usepackage{epsfig}
\usepackage{latexsym}

\newcommand{\postscript}[2]{\setlength{\epsfxsize}{#2\hsize}
   \centerline{\epsfbox{#1}}}
\newcommand{\comment}[1]{}

\def\MET{\mbox{${\hbox{$E$\kern-0.6em\lower-.1ex\hbox{/}}}_T$}}

\def\comment#1{{}}

\def\slashchar#1{\setbox0=\hbox{$#1$}           
   \dimen0=\wd0                                 
   \setbox1=\hbox{/} \dimen1=\wd1               
   \ifdim\dimen0>\dimen1                        
      \rlap{\hbox to \dimen0{\hfil/\hfil}}      
      #1                                        
   \else                                        
      \rlap{\hbox to \dimen1{\hfil$#1$\hfil}}   
      /                                         
   \fi}
\newif\ifnref

\nreffalse

\input epsf
\def\figin{\epsfcheck\figin}\def\figins{\epsfcheck\figins}
\def\epsfcheck{\ifx\epsfbox\UnDeFiNeD
\message{(NO epsf.tex, FIGURES WILL BE IGNORED)}
\gdef\figin##1{\vskip2in}\gdef\figins##1{\hskip.5in}
\else\message{(FIGURES WILL BE INCLUDED)}%
\gdef\figin##1{##1}\gdef\figins##1{##1}\fi}
\def\DefWarn#1{}
\def\figinsert{\goodbreak\midinsert}
\def\ifig#1#2#3{\DefWarn#1\xdef#1{fig.~\the\figno}
\writedef{#1\leftbracket fig.\noexpand~\the\figno}%
\figinsert\figin{\centerline{#3}}\medskip\centerline{\vbox{\baselineskip12pt
\advance\hsize by -1truein\noindent\footnotefont{\bf Fig.~\the\figno } #2}}
\bigskip\endinsert\global\advance\figno by1}

\def\hat{\widehat}

\begin{document}

\preprint{
\hfil
\begin{minipage}[t]{3in}
\begin{flushright}
\vspace*{.4in}
MPP--2009--159\\
LMU-ASC 39/09\\
CERN-PH-TH/2009-167
\vspace*{.4in}
\end{flushright}
\end{minipage}
}

\vspace{1cm}
\title{String Phenomenology at the LHC
\PRE{\vspace*{0.3in}} }

\author{Luis A. Anchordoqui}
\affiliation{Department of Physics,\\
University of Wisconsin-Milwaukee,
 Milwaukee, WI 53201, USA
\PRE{\vspace*{.1in}}
}

\author{Haim Goldberg}
\affiliation{Department of Physics,\\
Northeastern University, Boston, MA 02115, USA
\PRE{\vspace*{.1in}}
}

\author{Dieter \nolinebreak L\"ust}
\affiliation{Max--Planck--Institut f\"ur Physik\\
 Werner--Heisenberg--Institut,
80805 M\"unchen, Germany
\PRE{\vspace*{.1in}}
}

\affiliation{Arnold Sommerfeld Center for Theoretical Physics\\
Ludwig-Maximilians-Universit\"at M\"unchen,
80333 M\"unchen, Germany
\PRE{\vspace{.1in}}
}

\author{Stephan Stieberger}
\affiliation{Max--Planck--Institut f\"ur Physik\\
 Werner--Heisenberg--Institut,
80805 M\"unchen, Germany
\PRE{\vspace*{.1in}}
}

\author{Tomasz R. Taylor}
\affiliation{Department of Physics,\\
  Northeastern University, Boston, MA 02115, USA \PRE{\vspace*{.1in}}
}

\affiliation{Department of Physics, CERN Theory Division CH-1211 Geneva 23, Switzerland
\PRE{\vspace{.1in}}
}

\date{September 2009}
\PRE{\vspace*{.5in}}
\begin{abstract}\vskip 3mm
  \noindent We consider extensions of the standard model based on open strings
  ending on D-branes, with gauge bosons due to strings attached to
  stacks of D-branes and chiral matter due to strings stretching
  between intersecting D-branes.  Assuming that the fundamental string
  mass scale is in the TeV range and the theory is weakly coupled, we
  review possible signals of string physics at the Large Hadron
  Collider. 
\end{abstract}

\maketitle

At the time of its formulation and for years thereafter, superstring
theory was regarded as a unifying framework for Planck-scale quantum
gravity and TeV-scale standard model (SM) physics. Important advances
were fueled by the realization of the vital role played by
D-branes~\cite{joe} in connecting string theory to
phenomenology~\cite{Blumenhagen:2006ci}.  This has permitted the
formulation of string theories with compositeness setting in at TeV
scales and large extra dimensions~\cite{Antoniadis:1998ig}.

TeV-scale superstring theory provides a brane-world description of the
SM, which is localized on membranes extending in $p+3$ spatial
dimensions (here taken to be flat), the so-called D-branes. Gauge
interactions emerge as excitations of open strings with endpoints
attached on the D-branes, whereas gravitational interactions are
described by closed strings that can propagate in all nine spatial
dimensions of string theory (these comprise flat parallel
dimensions extended along the $(p+3)$-branes and transverse
dimensions). The apparent weakness of gravity at energies below few
TeVs can then be understood as a consequence of the gravitational
force ``leaking'' into the transverse compact dimensions of spacetime.

There are two paramount phenomenological consequences for TeV scale
D-brane string physics: the emergence of Regge recurrences at parton
collision energies $\sqrt{\hat s} \sim {\rm string\ scale} \equiv
M_s;$ and the presence of one or more additional $U(1)$ gauge
symmetries, beyond the $U(1)_Y$ of the SM. The latter follows from the
property that the gauge group for open strings terminating on a stack
of $N$ identical D-branes is $U(N)$ rather than $SU(N)$ for
$N>2$.\footnote{For $N=2$ the gauge group can be $Sp(1)$ rather than
  $U(2)$. The simplectic representation of Weinberg-Salam $SU(2)$
  reduces the required number of Higgs doublets to generate all Yukawa
  couplings at tree level~\cite{Cremades:2003qj}.} In this Brief
Review we exploit both these properties in order to identify ``new
physics'' signals at the Large Hadron Collider (LHC).

Only one assumption is necessary in order to set up a solid framework:
the string coupling must be small in order to rely on perturbation
theory in the computations of scattering amplitudes. In this case,
black hole production and other strong gravity effects occur at
energies above the string scale; therefore at least a few lowest Regge
recurrences are available for examination, free from interference with
some complex quantum gravitational phenomena.  Starting from a small
string coupling, the values of standard model coupling constants are
determined by D-brane configurations and the properties of extra
dimensions, hence that part of superstring theory requires intricate
model-building; however, as argued
in~\cite{Anchordoqui:2007da,Anchordoqui:2008ac,Anchordoqui:2008hi,Lust:2008qc,Anchordoqui:2008di,Anchordoqui:2009mm,Lust:2009pz},
some basic properties of Regge resonances like their production rates
and decay widths are completely model-independent.

To develop our program in the simplest way, we will work within the
construct of a minimal model in which we consider scattering processes
which take place on the (color) $U(3)$ stack of D-branes. In the
bosonic sector, the open strings terminating on this stack contain, in
addition to the $SU(3)$ octet of gluons $g_\mu^a$, an extra $U(1)$
boson ($C_\mu$, in the notation of~\cite{Berenstein:2006pk}), most
simply the manifestation of a gauged baryon number symmetry. The
$U(1)_Y$ boson $Y_\mu$, which gauges the usual electroweak hypercharge
symmetry, is a linear combination of $C_\mu$, the $U(1)$ boson $B_\mu$
terminating on a separate $U(1)$ brane, and perhaps a third additional
$U(1)$ sharing a $U(2)$ brane which is also a terminus for the
$SU(2)_L$ electroweak gauge bosons
$W_\mu^a$~\cite{Antoniadis:2000ena}. Any vector boson $Z_\mu'$,
orthogonal to the hypercharge, must grow a mass $M_{Z'}$ in order to
avoid long range forces between baryons other than gravity and Coulomb
forces. The anomalous mass growth allows the survival of global baryon
number conservation, preventing fast proton
decay~\cite{Ghilencea:2002da}. In what follows, the first Regge excitations of
the gluon, quarks, and the extra $U(1)$ boson tied to the color stack 
will be denoted by $g^*,\ q^*$, and $C^*$, respectively.

The physical processes underlying dijet production at the LHC are the
collisions of two partons $ij$, producing two final partons $kl$ that
fragment into hadronic jets. The corresponding $2\to 2$ scattering
amplitudes ${\cal M}(ij \to kl)$, computed at the leading order in
string perturbation theory, are collected in~\cite{Lust:2008qc}. The
amplitudes involving four gluons as well as those with two gluons plus
two quarks do not depend on the compactification details of the
transverse space.\footnote{The only remnant of the compactification is
  the relation between the Yang-Mills coupling and the string
  coupling. We take this relation to reduce to field theoretical
  results in the case where they exist, e.g., $gg \to gg$. Then,
  because of the require correspondence with field theory, the
  phenomenological results are independent of the compactification of
  the transverse space. However, a different phenomenology would
  result as a consequence of warping one or more parallel
  dimensions~\cite{Hassanain:2009at,Perelstein:2009qi}.}  All string
effects are encapsulated in these amplitudes in one ``form factor''
function of Mandelstam variables $\hat s,~\hat t,~\hat u$ (constrained
by $\hat s+\hat t+\hat u=0$)
\begin{equation}
V(  \hat s,   \hat t,   \hat u)= \frac{\hat s\,\hat u}{\hat tM_s^2}B(-\hat s/M_s^2,-\hat u/M_s^2)={\Gamma(1-   \hat s/M_s^2)\ \Gamma(1-   \hat u/M_s^2)\over
    \Gamma(1+   \hat t/M_s^2)}.\label{formf}
\end{equation}
The physical content of the form factor becomes clear after using the
well-known expansion in terms of $s$-channel resonances~\cite{Veneziano:1968yb}:
\begin{equation}
B(-\hat s/M_s^2,-\hat u/M_s^2)=-\sum_{n=0}^{\infty}\frac{M_s^{2-2n}}{n!}\frac{1}{\hat s-nM_s^2}
\Bigg[\prod_{J=1}^n(\hat u+M^2_sJ)\Bigg],\label{bexp}
\end{equation}
which exhibits $s$-channel poles associated to the propagation of
virtual Regge excitations with masses $\sqrt{n}M_s$. Thus near the
$n$th level pole $(\hat s\to nM^2_s)$:
\begin{equation}\qquad
V(  \hat s,   \hat t,   \hat u)\approx \frac{1}{\hat s-nM^2_s}\times\frac{M_s^{2-2n}}{(n-1)!}\prod_{J=0}^{n-1}(\hat u+M^2_sJ)\ .\label{nthpole}
\end{equation}
In specific amplitudes, the residues combine with the remaining
kinematic factors, reflecting the spin content of particles exchanged
in the $s$-channel, ranging from $J=0$ to $J=n+1$. 

The amplitudes for the four-fermion processes like quark-antiquark
scattering are more complicated because the respective form factors
describe not only the exchanges of Regge states but also of heavy
Kaluza-Klein (KK) and winding states with a model-dependent spectrum
determined by the geometry of extra dimensions. Fortunately, they are
suppressed, for two reasons. First, the QCD $SU(3)$ color group
factors favor gluons over quarks in the initial state. Second, the
parton luminosities in proton-proton collisions at the LHC, at the
parton center of mass energies above~1 TeV, are significantly lower
for quark-antiquark subprocesses than for gluon-gluon and
gluon-quark~\cite{Anchordoqui:2008ac}. The collisions of valence
quarks occur at higher luminosity; however, there are no Regge
recurrences appearing in the $s$-channel of quark-quark
scattering~\cite{Lust:2008qc}.

In the following we isolate the contribution to the partonic cross
section from the first resonant state. Note that far below the string
threshold, at partonic center of mass energies $\sqrt{\hat s}\ll M_s$,
the form factor $V(\hat s,\hat t,\hat u)\approx 1-\frac{\pi^2}{6}{\hat
  s \hat u}/M^4_s$~\cite{Lust:2008qc} and therefore the contributions
of Regge excitations are strongly suppressed. The $s$-channel pole
terms of the average square amplitudes contributing to dijet
production at the LHC can be obtained from the general formulae given
in~\cite{Lust:2008qc}, using Eq.(\ref{nthpole}). However, for
phenomenological purposes, the poles need to be softened to a
Breit-Wigner form by obtaining and utilizing the correct {\em total}
widths of the resonances~\cite{Anchordoqui:2008hi}. After this is
done, the contributions of the various channels to the spin and color
averaged matrix elements are as follows~\cite{Anchordoqui:2008di}
\begin{eqnarray}
|{\cal M} (gg \to gg)| ^2 & = & \frac{19}{12} \
\frac{g^4}{M_s^4} \left\{ W_{g^*}^{gg \to gg} \, \left[\frac{M_s^8}{(  \hat s-M_s^2)^2
+ (\Gamma_{g^*}^{J=0}\ M_s)^2} \right. \right.
\left. +\frac{  \hat t^4+   \hat u^4}{(  \hat s-M_s^2)^2 + (\Gamma_{g^*}^{J=2}\ M_s)^2}\right] \nonumber \\
   & + &
W_{C^*}^{gg \to gg} \, \left. \left[\frac{M_s^8}{(  \hat s-M_s^2)^2 + (\Gamma_{C^*}^{J=0}\ M_s)^2} \right.
\left. +\frac{  \hat t^4+  \hat u^4}{(  \hat s-M_s^2)^2 + (\Gamma_{C^*}^{J=2}\ M_s)^2}\right] \right\},
\label{gggg2}
\end{eqnarray}
\begin{eqnarray}
|{\cal M} (gg \to q \bar q)|^2 & = & \frac{7}{24} \frac{g^4}{M_s^4}\ N_f\
\left [W_{g^*}^{gg \to q \bar q}\, \frac{  \hat u \,  \hat t \, (\hat u^2+   \hat t^2)}{(  \hat s-M_s^2)^2 + (\Gamma_{g^*}^{J=2}\ M_s)^2} \right. \nonumber \\
 & + &  \left. W_{C^*}^{gg \to q \bar q}\, \frac{  \hat u \,  \hat t \, (  \hat u^2+ \hat   t^2)}{( \hat s-M_s^2)^2 +
(\Gamma_{C^*}^{J=2}\ M_s)^2} \right]
\end{eqnarray}
\begin{eqnarray}
|{\cal M} (q \bar q \to gg)|^2  & = &  \frac{56}{27} \frac{g^4}{M_s^4}\
\left[ W_{g^*}^{q\bar q \to gg} \,  \frac{ \hat u \, \hat t \, ( \hat  u^2+ \hat  t^2)}{( \hat s-M_s^2)^2 + (\Gamma_{g^*}^{J=2}\ M_s)^2} \right. \nonumber \\
 & + & \left.  W_{C^*}^{q\bar q \to gg} \, \frac{ \hat u \, \hat t \,( \hat  u^2+ \hat  t^2)}{( \hat s-M_s^2)^2 + (\Gamma_{C^*}^{J=2}\ M_s)^2} \right] \,\,,
\end{eqnarray}
\begin{equation}
|{\cal M}(qg \to qg)|^2  =  - \frac{4}{9} \frac{g^4}{M_s^2}\
\left[ \frac{M_s^4 \hat  u}{( \hat s-M_s^2)^2 + (\Gamma_{q^*}^{J=1/2}\ M_s)^2} + \frac{\hat u^3}{(\hat s-M_s^2)^2 + (\Gamma_{q^*}^{J=3/2}\ M_s)^2}\right],
\label{qgqg2}
\end{equation}
where $g$ is the QCD coupling constant $(\alpha_{\rm QCD}=\frac{g^2}{4\pi}\approx 0.1)$
 and $\Gamma_{g^*}^{J=0} = 75\, (M_s/{\rm TeV})~{\rm GeV}$,
$\Gamma_{C^*}^{J=0} = 150 \, (M_s/{\rm TeV})~{\rm GeV}$,
$\Gamma_{g^*}^{J=2} = 45 \, (M_s/{\rm TeV})~{\rm GeV}$,
$\Gamma_{C^*}^{J=2} = 75 \, (M_s/{\rm TeV})~{\rm GeV}$,
$\Gamma_{q^*}^{J=1/2} = \Gamma_{q^*}^{J=3/2} = 37\, (M_s/{\rm
  TeV})~{\rm GeV}$ are the total decay widths for intermediate states
$g^*$, $C^*$, and $q^*$ (with angular momentum
$J$)~\cite{Anchordoqui:2008hi}. The associated weights of these
intermediate states are given in terms of the probabilities for the
various entrance and exit channels
\begin{equation}
W_{g^*}^{gg \to gg} = \frac{8(\Gamma_{g^* \to gg})^2}{8(\Gamma_{g^* \to gg})^2 +
(\Gamma_{C^* \to gg})^2} = 0.44 \,,
\label{w1}
\end{equation}
\begin{equation}
W_{C^*}^{gg \to gg} = \frac{(\Gamma_{C^*
  \to gg})^2}{8(\Gamma_{g^* \to gg})^2 + (\Gamma_{C^* \to gg})^2} =
0.56 \, ,
\label{w2}
\end{equation}
\begin{equation}
W_{g^*}^{gg \to q \bar q}  = W_{g^*}^{q \bar q \to gg} =
\frac{8\,\Gamma_{g^* \to gg} \,
\Gamma_{g^* \to q \bar q}} {8\,\Gamma_{g^* \to gg} \,
\Gamma_{g^* \to q \bar q} + \Gamma_{C^* \to gg} \,
\Gamma_{C^* \to q \bar q}} = 0.71 \, ,
\label{w3}
\end{equation}
\begin{equation}
W_{C^*}^{gg \to q \bar q} = W_{C^*}^{q \bar q \to gg}  =
\frac{\Gamma_{C^* \to gg} \,
\Gamma_{C^* \to q \bar q}}{8\,\Gamma_{g^* \to gg} \,
\Gamma_{g^* \to q \bar q} + \Gamma_{C^* \to gg} \,
\Gamma_{C^* \to q \bar q}} = 0.29 \, .
\label{w4}
\end{equation}
Superscripts $J=2$ are understood to be inserted on all the $\Gamma$'s in
Eqs.(\ref{w1}), (\ref{w2}), (\ref{w3}), (\ref{w4}).
Equation~(\ref{gggg2}) reflects the fact that weights for $J=0$ and
$J=2$ are the same~\cite{Anchordoqui:2008hi}. In what follows we set
the number of flavors $N_f =6$.

Next, we obtain the dominant $s$-channel pole terms of the average square
amplitudes contributing to $pp \to \gamma$ + jet~\cite{Anchordoqui:2007da}
\begin{eqnarray}
|{\cal M}(qg \to q \gamma)|^2   =   -\frac{1}{3} Q^2 \frac{g^4}{M_s^2}\
\left[ \frac{M_s^4  \hat  u}{( \hat s-M_s^2)^2 + (\Gamma_{q^*}^{J=\frac{1}{2}}\ M_s)^2}  \right.  + \left. \frac{\hat u^3}{(\hat s-M_s^2)^2 + (\Gamma_{q^*}^{J=\frac{3}{2}}\ M_s)^2}\right]
\label{qgqz}
\end{eqnarray}
and
\begin{eqnarray}
|{\cal M} (gg \to g \gamma)|^2  =  \frac{5}{3} \, Q^2 \,
\frac{g^4}{M_s^4}  \, \left[\frac{M_s^8}{( \hat s-M_s^2)^2
+ (\Gamma_{g^*}^{J=0}\ M_s)^2}  + \frac{  \hat t^4+  \hat u^4}{( \hat s-M_s^2)^2 + (\Gamma_{g^*}^{J=2}\ M_s)^2}\right] \,,
\label{gggz}
\end{eqnarray}
where $Q = \sqrt{1/6} \ \kappa \ \cos \theta_W$ is the product of the
$U(1)$ charge of the fundamental representation ($\sqrt{1/6}$)
followed by successive projections onto the hypercharge ($\kappa$) and
then onto the photon ($\cos \theta_W$).  The $C-Y$ mixing coefficient
is model dependent: in the minimal $U(3) \times Sp(1) \times U(1)$
model it is quite small, around $\kappa \simeq 0.12$ for couplings
evaluated at the $Z$ mass, which is modestly enhanced to $\kappa \
\simeq 0.14$ as a result of RG running of the couplings up to 2.5~TeV.
It should be noted that in models possessing an additional $U(1)$
which partners $SU(2)_L$ on a $U(2)$ brane, the various assignment of
the charges can result in values of $\kappa$ which can differ
considerably from $0.12.$

Before proceeding we pause to stress that at low energies 
\begin{equation}
\label{mhvlow}
|{\cal M}(gg\to  g\gamma)|^2\approx  g^4Q^2C(N){\pi^4\over 4}
(\hat s^4+\hat t^4+\hat u^4)\qquad
  (\hat s,\hat t,\hat u\ll 1) \, .
\end{equation}
The absence of massless poles, at $s=0$ {\it etc.\/}, translated
into the terms of effective field theory, confirms that there are
no exchanges of massless particles contributing to this process.

Events with a single jet plus missing energy ($\MET$) with balancing
transverse momenta (so-called ``monojets'') are incisive probes of new
physics. As in the SM, the source of this topology is $ij \to k Z^0$
followed by $Z^0 \to \nu \bar \nu.$ Both in the SM and string theory
the cross section for this process is of order $g^4$. Virtual KK
graviton emission ($ij \to k G$) involves emission of closed strings,
resulting in an additional suppression of order $g^2$ compared to
$Z^0$ emission. A careful discussion of this suppression is given
in~\cite{Cullen:2000ef}. However, in some scenarios compensation for
this suppression can arise from the large multiplicity of graviton
emission, which is somewhat dependent on the cutoff
mechanism~\cite{Bando:1999di,Anchordoqui:2001cg,Hewett:2007st}. Ignoring
the $Z$-mass (i.e., keeping only transverse $Z$'s), the quiver
contribution to $pp \to Z + {\rm jet}$ is suppressed relative to the
$pp \to \gamma+ {\rm jet}$ by a factor of $\tan^2\theta_W = 0.29.$

\begin{figure}[tbp]
\begin{minipage}[t]{0.32\textwidth}
\postscript{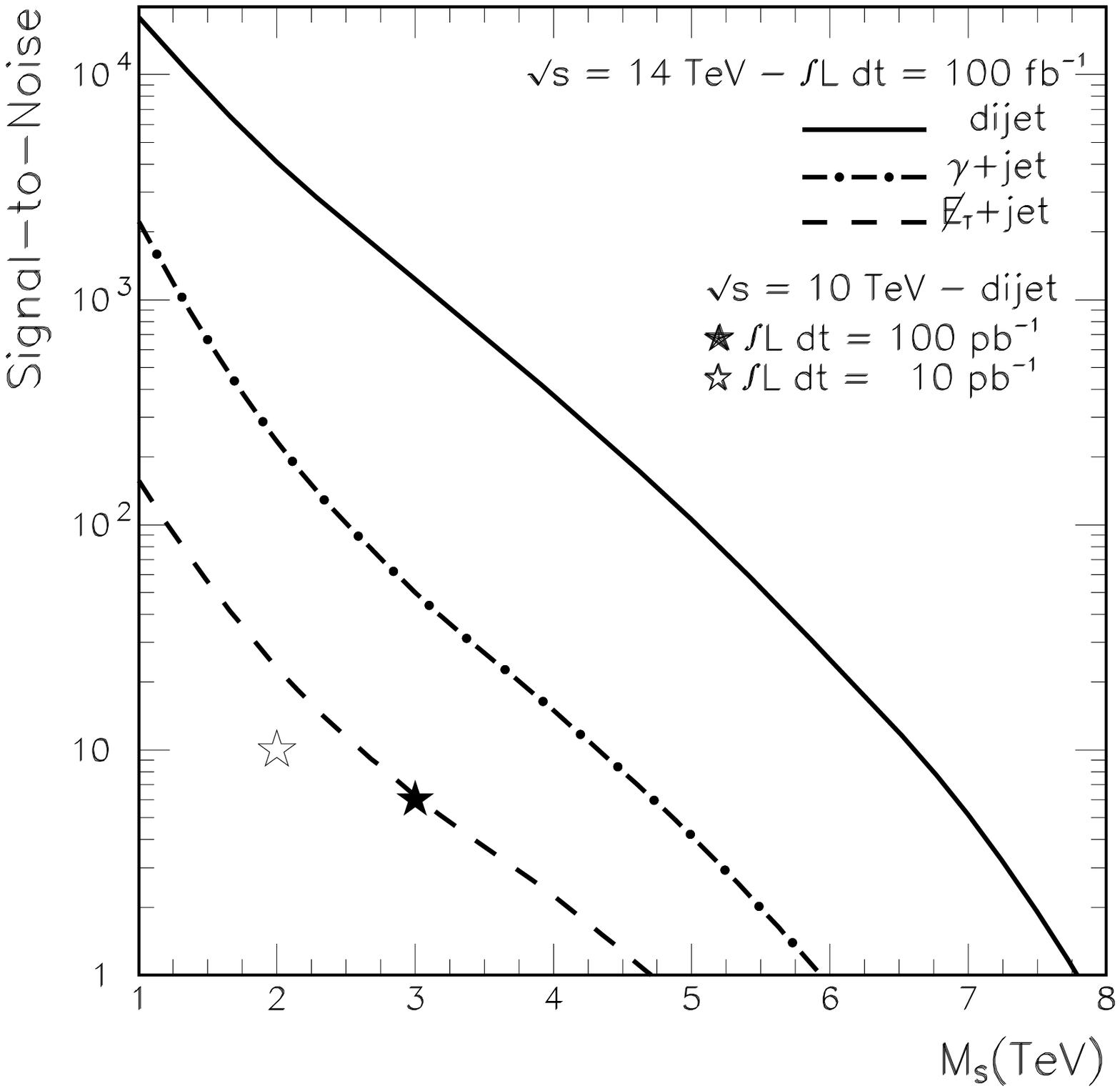}{0.99}
\end{minipage}
\begin{minipage}[t]{0.32\textwidth}
\postscript{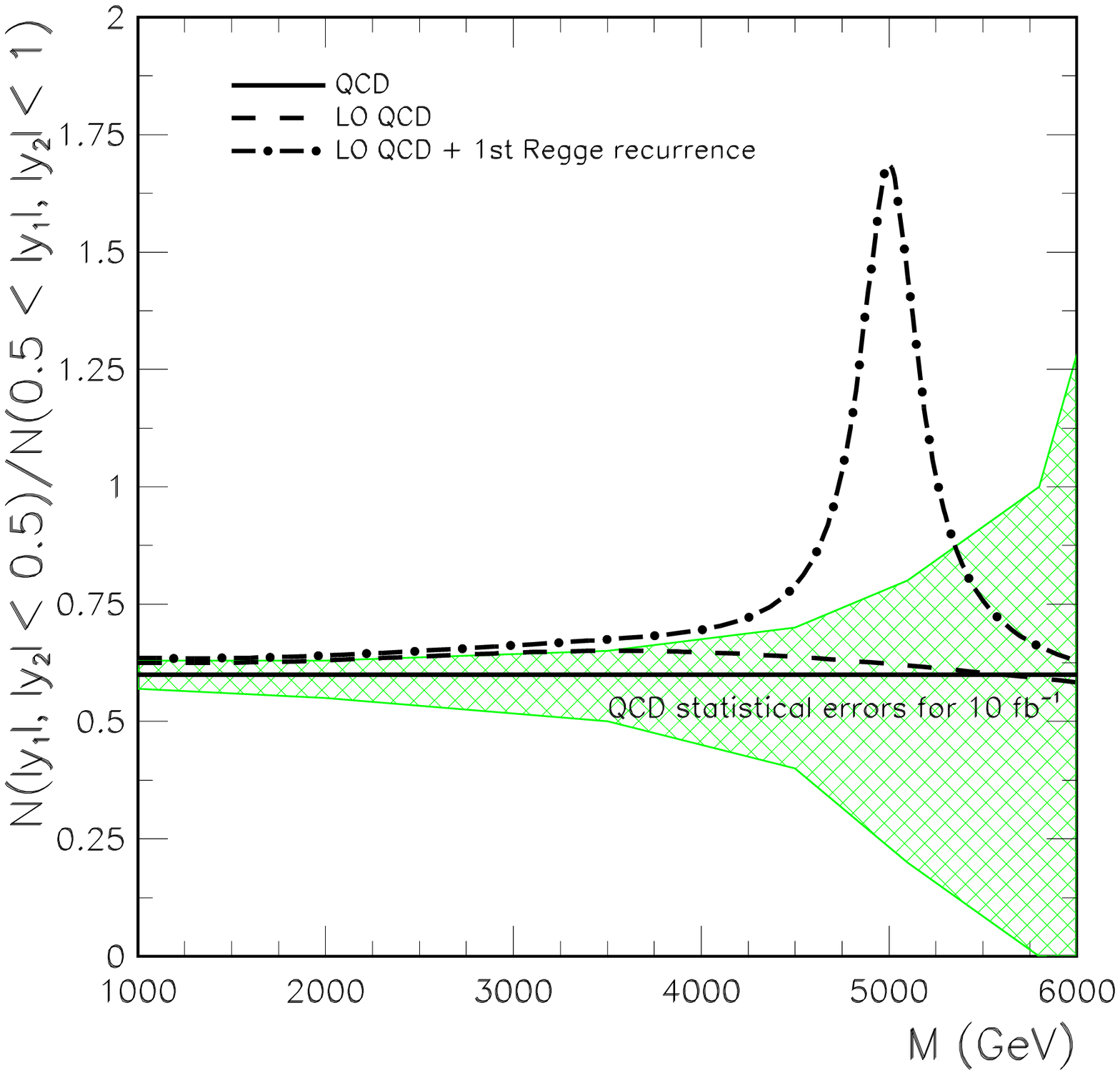}{0.99}
\end{minipage}
\begin{minipage}[t]{0.32\textwidth}
\postscript{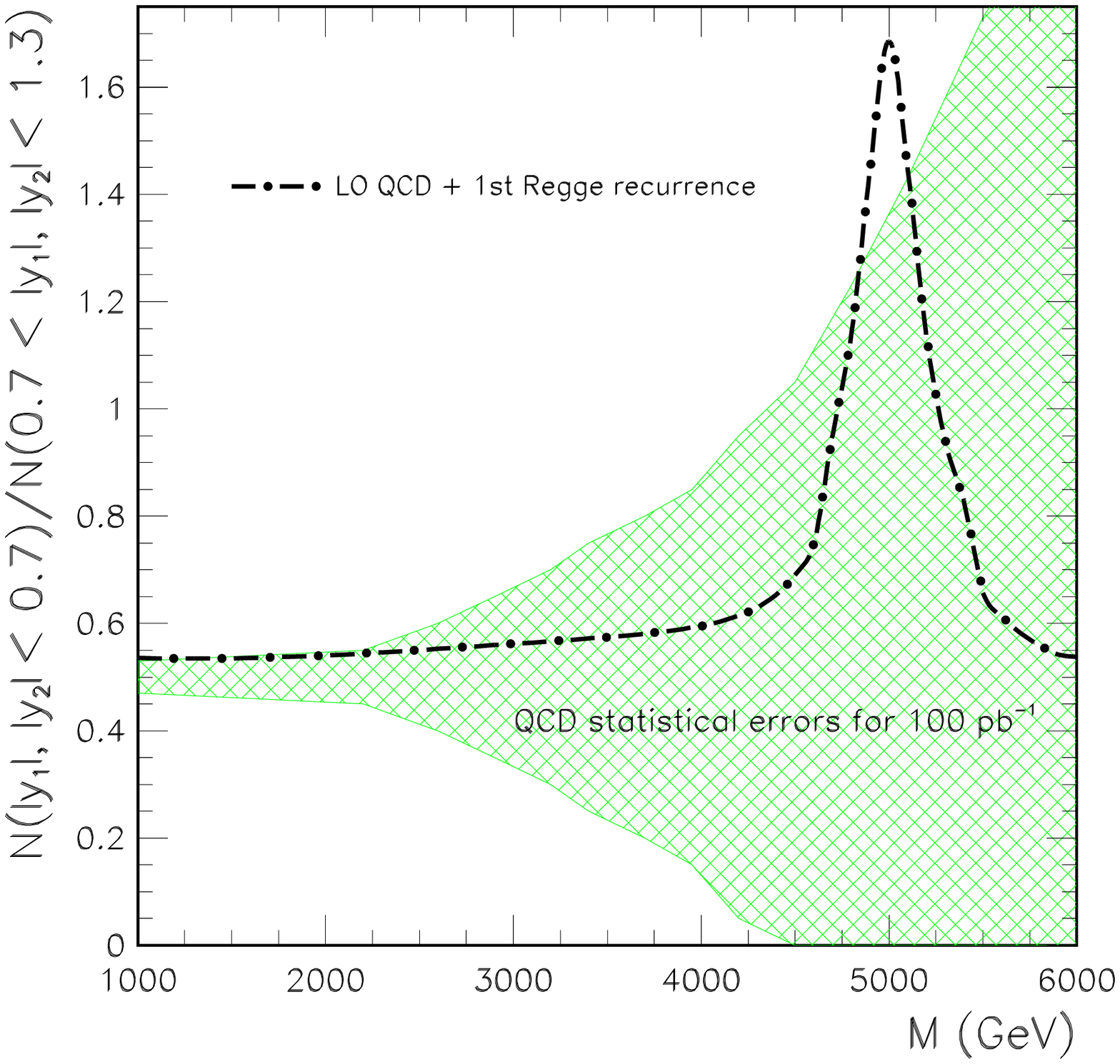}{0.99}
\end{minipage}
\caption{{\em Left panel:} Signal-to-noise ratio of $pp \to {\rm
    dijet}$, $pp \to \gamma + {\rm jet}$, and $pp \to \MET + {\rm
    jet}$, for $\sqrt{s} = 14$~TeV, $\sqrt{s} = 10$~TeV, $\kappa^2
  \simeq 0.02$, and various integrated luminosities. The approximate
  equality of the background due to misidentified $\pi^0$'s and the
  QCD background, across a range of large $p_T^\gamma$ as implemented
  in~\cite{Anchordoqui:2008ac}, is maintained as an approximate
  equality over a range of invariant $\gamma$-jet invariant masses
  with the rapidity cuts imposed. The monojet signal is obtained from
  the intermediate state $pp \to Z^0+$ jet multiplied by the
  corresponding branching ratio $Z^0 \to \nu\bar \nu$.  {\em Middle
    panel:} For a luminosity of 10~fb$^{-1}$ and $\sqrt{s} = 14$~TeV,
  the expected value (solid line) and statistical error (shaded
  region) of the dijet ratio of QCD in the CMS detector is compared
  with LO QCD (dashed line) and LO QCD plus lowest massive string
  excitation (dot-dashed line), at a scale $M_s = 5$~TeV. {\em Right
    panel:} The QCD statistical errors of the new dijet ratio are
  compared with the predictions for LO QCD plus the lowest massive string
  excitation (dot-dashed line) at $\sqrt{s} = 14$~TeV, for
  100~pb$^{-1}$.}
\label{figure}
\end{figure}

The first Regge recurrence would be visible in data binned according
to the invariant mass $M$ of the final state, after setting cuts on
rapidities $|y_1|, \, |y_2| \le y_{\rm max}$ and transverse momenta
$p_{\rm T}^{1,2}>50$~GeV, where $y_{\rm max} = 2.4$ for photons and
$y_{\rm max} = 1$ for jets. The QCD background is calculated at the
partonic level making use of the CTEQ6D parton distribution
functions~\cite{Pumplin:2002vw}. Standard bump-hunting methods, such
as obtaining cumulative cross sections, $\sigma (M_0 ) =
\int_{M_0}^\infty \frac{d\sigma}{dM}\, dM$, from the data and
searching for regions with significant deviations from the QCD
background, may reveal an interval of $M$ suspected of containing a
bump. With the establishment of such a region, one may calculate a
signal-to-noise ratio, with the signal rate estimated in the invariant
mass window $[M_s - 2\Gamma, M_s + 2\Gamma].$ The noise is defined as
the square root of the number of background events in the same dijet
mass interval for the same integrated luminosity. The LHC discovery
reach (at the parton level) is encapsulated in Fig.~\ref{figure}. The
solid, dot-dashed, and dashed lines show the behavior of the
signal-to-noise (S$/\sqrt{\rm B}$) ratio as a function of the string
scale for three different event topologies (dijet, $\gamma +$ jet, and
$\MET +$ jet; respectively), at $\sqrt{s} = 14$~TeV with an integrated
luminosity of 100~fb$^{-1}$. It is remarkable that with 100~fb$^{-1}$
of data collection, {\it string scales as large as 6.8~TeV are open to
  discovery at the ≥ $5\sigma$ level.} Although the discovery reach is
not as high as that for dijets, the measurement of
$pp\rightarrow\gamma~ +$ jet and $pp \to \MET +$ jet can potentially
provide an interesting corroboration for the stringy origin for new
physics manifest as a resonant structure in LHC data. The stars in 
Fig.~\ref{figure} show the expected S/N of dijet events for the first
LHC run, at $\sqrt{s} = 10$~TeV. For $M_s = 3$~TeV and 100~pb$^{-1}$
of data collected at $\sqrt{s} = 10$~TeV, we expect S$/\sqrt{\rm B} =
127/20 = 6 \sigma.$ For an overly conservative assumption of
integrated luminosity $\approx$ 10~pb$^{-1}$, a S$/\sqrt{\rm B} =
204/19 > 10\sigma$ is expected for string scales as high as $M_s =
2$~TeV. It is also remarkable that within 1 year of data collection at
$\sqrt{s} = 10$~TeV, {\it string scales as large as 3~TeV are open to
  discovery at the $\geq 5 \sigma$ level.} Once more, we stress that
these results contain no unknown parameters. They depend only on the
D-brane construct for the SM, and {\it are independent of
  compactification details of the transverse space.}

We now turn to the analysis of the angular distributions. QCD
parton-parton cross sections are dominated by $t$-channel exchanges
that produce dijet angular distributions which peak at small center of
mass scattering angles. In contrast, non--standard contact
interactions or excitations of resonances result in a more isotropic
distribution. In terms of rapidity variables for standard transverse
momentum cuts, dijets resulting from QCD processes will preferentially
populate the large rapidity region, while the new processes generate
events more uniformly distributed in the entire rapidity region. To
analyze the details of the rapidity space the D\O\
Collaboration~\cite{Abbott:1998wh} introduced a new parameter $R$, the
ratio of the number of events, in a given dijet mass bin, for both
rapidities $|y_1|, |y_2| < 0.5$ and both rapidities $0.5 < |y_1|,
|y_2| < 1.0$.\footnote{An illustration of the use of this parameter in
  a heuristic model where standard model amplitudes are modified by a
  Veneziano formfactor has been presented ~\cite{Meade:2007sz}.}  In
Fig.~\ref{figure} we compare the results from a full CMS detector
simulation of the ratio $R$~\cite{Esen}, with predictions from LO QCD
and model-independent contributions to the $q^*$, $g^*$ and $C^*$
excitations~\cite{Anchordoqui:2008di}.  For an integrated luminosity
of 10~fb$^{-1}$ the LO QCD contributions with $\alpha_{\rm QCD} = 0.1$
(corresponding to running scale $\mu \approx M_s$) are within
statistical fluctuations of the full CMS detector simulation. (Note
that the string scale is an optimal choice of the running scale which
should normally minimize the role of higher loop corrections.) Since
one of the purposes of utilizing NLO calculations is to fix the choice
of the running coupling, we take this agreement as rationale to omit
loops in QCD and in string theory. It is clear from Fig.~\ref{figure}
that incorporating NLO calculation of the background and the signal
would not significantly change the large deviation of the string
contribution from the QCD background. Very recently the CMS
Collaboration optimized the dijet ratio~\cite{Bhatti:2008hz}. The new
ratio, $N(|y_1|, |y_2| < 0.7)/N(0.7 < |y_1|, |y_2| <1.3),$ is shown in
Fig.~\ref{figure}; string scales $M_s < 5$~TeV can be probed with
100~pb$^{-1}$ of data collection.

Although there are no $s$-channel resonances in $qq\rightarrow qq$ and
$qq'\rightarrow qq'$ scattering, KK modes in the $t$ and $u$ channels
generate calculable effective 4-fermion contact
terms~\cite{Lust:2008qc}. These in turn are manifest in an enhancement
in the continuum below the string scale of the $R$ ratio for dijet
events. For $M_{\rm KK}\le 3$~TeV, this contribution can be detected
at the LHC with 6$\sigma$ significance above SM
background~\cite{Anchordoqui:2009mm}.  In combination with the
simultaneous observation in dijet events of a string resonance at
$M_s> M_{\rm KK}$, this would consolidate the stringy interpretation
of these anomalies. In particular, it could serve to differentiate
between a stringy origin for the resonance as opposed to an isolated
structure such as a $Z'$, which would not modify $R$ outside the
resonant region. Moreover, because of the high multiplicity of the
angular momenta (up to $J=2$), the rapidity distribution of the decay
products of string excitations would differ significantly from those
following decay of a $Z'$ with $J=1$.  With high statistics, isolation
of lowest massive Regge excitations from KK replicas (with $J=2$) may
also be possible.

In terms of the perturbative cross section itself, the multiplicity of
levels that one excites is limited because the scattering takes place
on a particular brane from a particular helicity state. For level $n$,
the Regge trajectory has access to $n$ spins. For fixed $J$ the
coupling of the heavier string resonance modes $\alpha_J^n$ decreases
quadratically with energy~\cite{Cornet:2001gy}, but the number of
modes at each mass level $n$ grows also quadratically. This makes
$\sum_J \alpha_n^J$ a constant independent of $n$ and so the cross
section grows linearly with energy. However, the D-brane structure
with scattering confined to a single stack of D-branes may not be a
valid assumption for the decay of higher mass level excitated states,
because there may be branching into a high multiplicity of two
resonances which connect to different branes.  Generally, the width of
the Regge excitations will grow at least linearly with energy, whereas
the spacing between levels will decrease with energy. This implies an
upper limit on the domain of validity for this phenomenological
approach. In particular, for a resonance $R$ of mass $M$, the total
width $\Gamma_{\rm tot} =  \alpha_{\rm QCD} \, {\cal C} M/4$, where
${\cal C} > 1$ because of the growing multiplicity of decay modes.  On
the other hand, the level spacing at mass $M$ is $\Delta M =
M_s^2/M$. Therefore,
\begin{equation}
\frac{\Gamma_{\rm tot}}{\Delta M} = \frac{g^2}{16 \pi} \ {\cal C} \ \left(\frac{M}{M_s} \right)^2 = \frac{g^2}{16 \pi} \ {\cal C} \ n <1 \, .
\end{equation}
For excitation of the resonance $R$ via $a+b\rightarrow R$, the
assumption $\Gamma_{\rm tot}(R) \sim \Gamma(R\rightarrow ab)$ (which
underestimates the real width) yields a perturbative regime for $n <
50$~\cite{Cornet:2001gy}. For an increase width, the transition level
can easily drop to $n=10$.

Black hole intermediate states are expected to dominate $s$-channel
scattering at trans-Planckian
energies~\cite{Amati:1987wq,'tHooft:1987rb,Amati:1992zb,Amati:2007ak}. The
number of such non-perturbative states grows faster than that of any
perturbative state; e.g., the number of black hole states in 10
spacetime dimensions grows with mass like $e^{M^{8/7}}$, whereas the
number of perturbative string states grows like $e^M$. Along these
lines, if $M_s \simeq 1~{\rm TeV}$, semiclassical arguments seem to
indicate that black hole production and evaporation would be observed
at the LHC~\cite{Dimopoulos:2001hw,Giddings:2001bu}. However, for $M_s
\agt 2~{\rm TeV}$, the LHC black holes would become stringy
(a.k.a. ``string balls'') and their properties rather complex. String
ball dynamics is framed on the context of the string
$\rightleftharpoons$ black hole correspondence principle: when the
size of the black hole horizon drops below the size of the fundamental
string length, an adiabatic transition occurs to an excited string
state~\cite{Horowitz:1996nw,Damour:1999aw}. Subsequently, the string
will slowly lose mass by radiating massless particles with a nearly
thermal spectrum at the unchanging Hagedorn
temperature~\cite{Amati:1999fv}.\footnote{The probability that a black
  hole will radiate a large string, or else that a large string would
  undergo fluctuation to become a black hole is very
  small~\cite{Horowitz:1997jc}.}  The continuity of the cross section
at the correspondence point, at least parametrically in energy and
string coupling, provides independent supportive argument for this
picture~\cite{Dimopoulos:2001qe}.

{\em In summary,} in D-brane constructions, the full-fledged string
amplitudes supplying the dominant contributions to dijet cross
sections are completely independent of the details of
compactification. If the string scale is in the TeV range, such
extensions of the standard model can be of immediate phenomenological
interest. In this Brief Review we have made use of the amplitudes
evaluated near the first resonant pole to report on the discovery
potential at the LHC for the first Regge excitations of the quark and
gluon. Remarkably, after a few years of running, the reach of LHC in
the dijet topology ($S/N = 210/42$) can be as high as 6.8~TeV. This
intersects with the range of string scales consistent with correct
weak mixing angle found in the $U(3) \times U(2) \times U(1)$ quiver
model~\cite{Antoniadis:2000ena}. For string scales as high as 5.0 TeV,
observations of resonant structures in $pp\rightarrow \gamma~ +$ jet
can provide interesting corroboration for stringy physics at the
TeV-scale.

\section*{Acknowledgments}
L.A.A.\ is supported by the U.S. National Science Foundation Grant No
PHY-0757598, and the UWM Research Growth Initiative.  H.G.\ is
supported by the U.S. National Science Foundation Grant No
PHY-0757959.  The research of D.L.\ and St.St.\ are supported in part
by the European Commission under Project MRTN-CT-2004-005104. The
research of T.R.T.\ is supported by the U.S.  National Science
Foundation Grants PHY-0600304, PHY-0757959 and by the Cluster of
Excellence ``Origin and Structure of the Universe'' in Munich,
Germany. Any opinions, findings, and conclusions or recommendations
expressed in this material are those of the authors and do not
necessarily reflect the views of the National Science Foundation.

\end{document}